\documentclass[aps,preprint]{revtex4}
\usepackage{graphicx}
\usepackage{bm}
\usepackage{cases}
\usepackage{amssymb}
\usepackage{slashed}
\usepackage{color}

\usepackage{subfigure}

\begin{document}
\title{Proposal of A Linked Mirror Configuration for Magnetic Confinement Experiment}
\author{Zhichen Feng}
\author{Guodong Yu}
\author{Peiyou Jiang}
\author{GuoYong Fu}\thanks{corresponding author's Email: gyfu@zju.edu.cn}
\affiliation{\small{Institute for Fusion Theory and Simulation and Department of physics, Zhejiang University, Hangzhou 310027, China}}

\begin{abstract}

A new linked mirror device for magnetic confinement experiment is proposed. The new linked mirror device consists of two straight magnetic mirrors connected by two half-torus. The structure of the configuration as a whole is three dimensional because the two linear mirror sections are not parallel. The angle between the two mirror sections generates rotational transform which results in good magnetic confinement of toroidally processing passing particles. In this way the usual loss cone of the traditional linear mirror machines is eliminated. The single particle confinement is similar to that of tokamaks with most of particles well confined. The calculated neoclassical confinement is very good and is even better than that of an equivalent tokamak. Potentially the proposed linked mirror configuration can allow MHD stable high beta equilibria with good plasma confinement suitable for neutron sources and magnetic fusion reactors.

\end{abstract}

\baselineskip 18pt \textwidth6.5in\textheight9in

\maketitle

\section{introduction}

Magnetically controlled fusion energy research has been pursued since early 1950's. Many types of magnetic confinement devices have been invented, such as magnetic mirrors\cite{Fowler,Post}, stellarators\cite{Spitzer,Wakatani} and tokamaks. One of earliest and simplest magnetic confinement devices is magnetic mirror. In early days the magnetic mirror devices suffered from two major defects: the fast end loss and strong MHD interchange instability\cite{Post}. In particular the end loss of passing particles is very fast leading to poor plasma confinement. Since then various schemes have been developed to mitigate the fast end loss, such as tandem mirror approach\cite{Fowler2} and the more modern approach of GDT\cite{Burdakov,Ivanov}. These approaches have indeed improved the plasma confinement and MHD stability considerably. However some residual end loss remains and as a result plasma confinement is still not good enough for fusion reactors.

In this paper a new linear-torus hybrid device of Linked Mirror Configuration (LMC) has been proposed. The main idea of the new configuration was inspired by the concepts of linear magnetic mirror and the figure-8 stellarator\cite{Spitzer}. The hybrid configuration consists of two straight mirror sections connected by two half-torus sections. The two linear mirror sections are not parallel but with a finite angle between them. This angle results in three dimensional (3D) toroidal structure of magnetic field leading to finite rotational transform, just as in the early figure-8 stellarator. Correspondingly the toroidally transiting passing particles are well confined and the end loss of a linear mirror is eliminated. Indeed our simulation results show that the single particle confinement is very good and the neoclassical transport is very low, even lower than that of an equivalent tokamak.

The linked mirror approach has been considered previously. The earlier ELMO Bumpy Torus (EBT) consists of many mirrors linked toroidally\cite{EBT}. Later in 1981 the DRAKON closed fusion trap was proposed by Glagolev et al.\cite{Glagolev,Kondakov}, consists of two long linear mirror sections joined by two curvilinear sections (CRELs). More recently, Mazzucato proposed a fusion reactor scheme of two linear mirrors joined by two toroidal sections with helical coils\cite{Mazzucato}. Conceptually our linked mirror approach is similar to those of the DRAKON\cite{Glagolev,Kondakov} and Mazzucato's\cite{Mazzucato}. The main difference is that in our approach the rotational transform is generated by the angle between the two non-parallel mirror sections rather than the two 3D half-torus sections.

As will be shown later, our LMC posses following good properties: (1) there is no end loss; (2) neoclassical confinement is very good and better than that of equivalent tokamaks; (3)All the current coils have the simple geometry of circular planar shape. Finally the device is potentially MHD stable order-of-unity plasma beta. Because of these good properties, the proposed LMC is potentially suitable for neutron sources and fusion reactors.

The paper is organized as follows. Section II introduces the basic structure of the LMC. Section III describes the rotational transform properties. Section IV analyses the quasi-omnigeneity and single particle confinement of LMC. Section V presents the numerical results of neoclassical transport. Section VI briefly discusses MHD stability. Section VII gives conclusions.

\section{Basic structures of the linked mirror configuration}

\begin{figure}[h]
\subfigure[Top view of the linked mirror configuration]{
\label{fig1.b}
\includegraphics[scale=0.2]{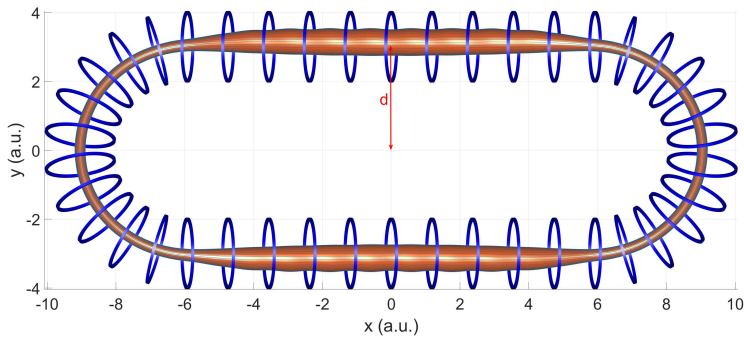}}
\subfigure[Side view of the linked mirror configuration]{
\label{fig1.c}
\includegraphics[scale=0.2]{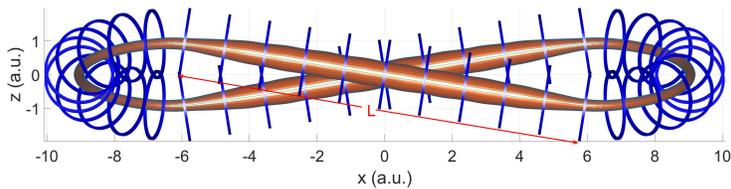}}
\caption{Basic structure of the linked mirror configuration}\label{fig1}
\end{figure}

The basic structure of LMC is shown in Fig. \ref{fig1} with cartesian coordinates (x,y,z). From the top view (Fig. \ref{fig1.b}),
the configuration shape looks like a race track. The configuration consists of two straight mirror sections and two half-torus sections.
However the two straight sections are not parallel as seen from the side view in Fig. \ref{fig1.c}.
The angle between the two straight sections is defined as $2\Theta$ with $\Theta$ being the angle between one of the straight section
and the horizontal plane (z=0 plane). We will show in next section that this angle is responsible for generating rotational transform of the whole configuration.
The magnetic field is provided by simple circular planar coils of equal radius and different currents.
For the two mirror sections, the coils forms a circular cylinder. On the other hand, for the half-torus sections,
the circular coils are placed in such a way that the centers of the coils form a half circle connecting the two ends of the mirror sections.
Obviously, the connections between the straight and half-torus sections can not be tangent, but the magnetic field lines and the magnetic flux surfaces are always smooth, as will be shown later. Besides the angle $\Theta$, other key configuration parameters include $L$ the length of the straight mirror section, $d$ the half distance between the two mirror sections, $r$ the radius of the coils, and $I_i$ the current in each coil.

Fig. \ref{fig1} plots the coils in blue, one of the magnetic surfaces in orange, and a magnetic field line on the flux surface. The parameters used are: $\Theta=\pi/20$, $L=12m$, $d=3m$, $r=1m$, $I_{torus}/I_{tube}=5I_0:1I_0$,
which means the currents in the coils of the half-torus sections are 5 times of that in the straight mirror tubes. It should be pointed out that the currents in the four end coils in each mirror section are $5I_0$ too.

\begin{figure}
\subfigure[Cross section of magnetic flux surfaces at $\phi= 0$ plane]{
\label{fig2.a}
\includegraphics[scale=0.17]{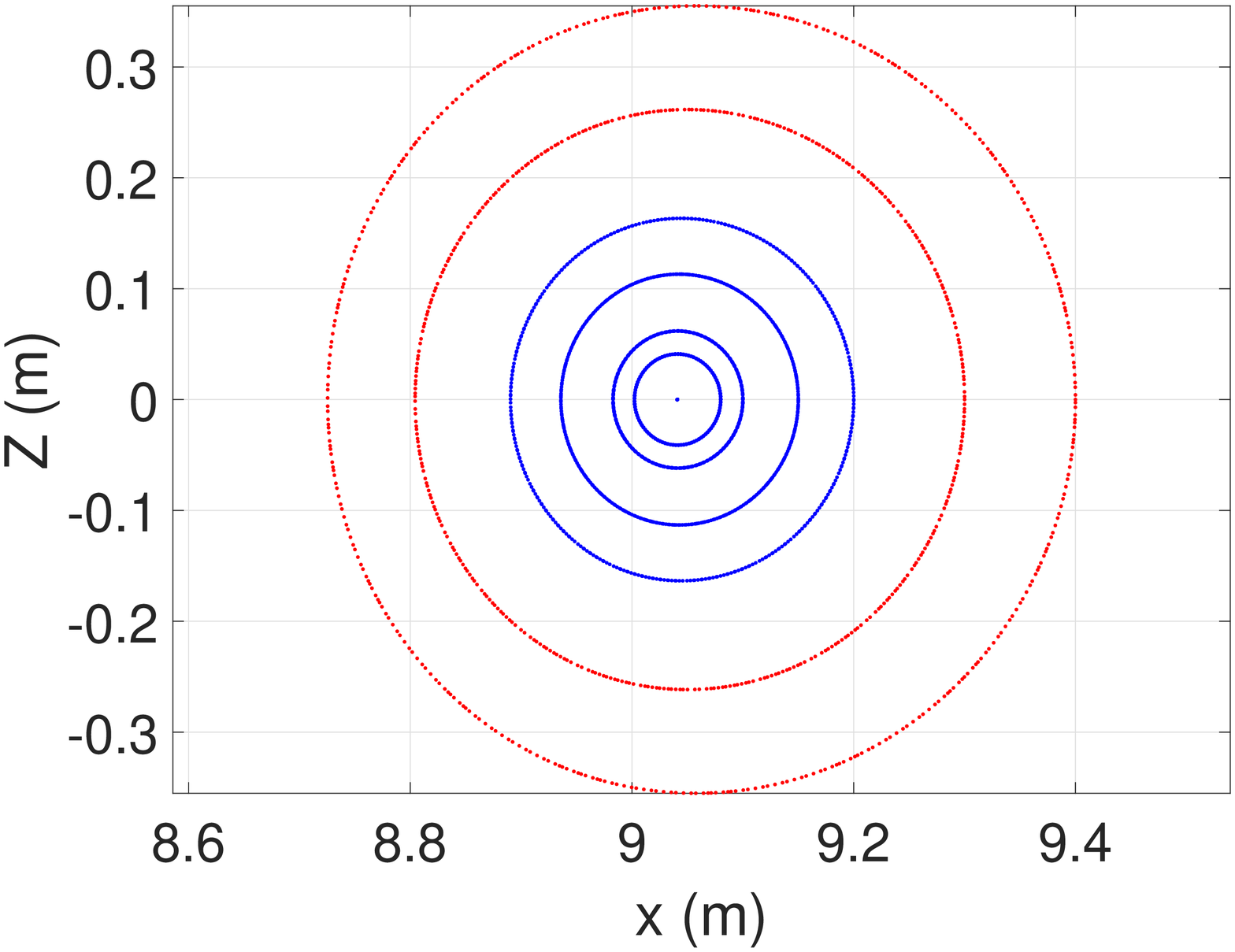}}
\subfigure[Cross section of magnetic flux surfaces at $\phi= \pi/2$ plane]{
\label{fig2.b}
\includegraphics[scale=0.25]{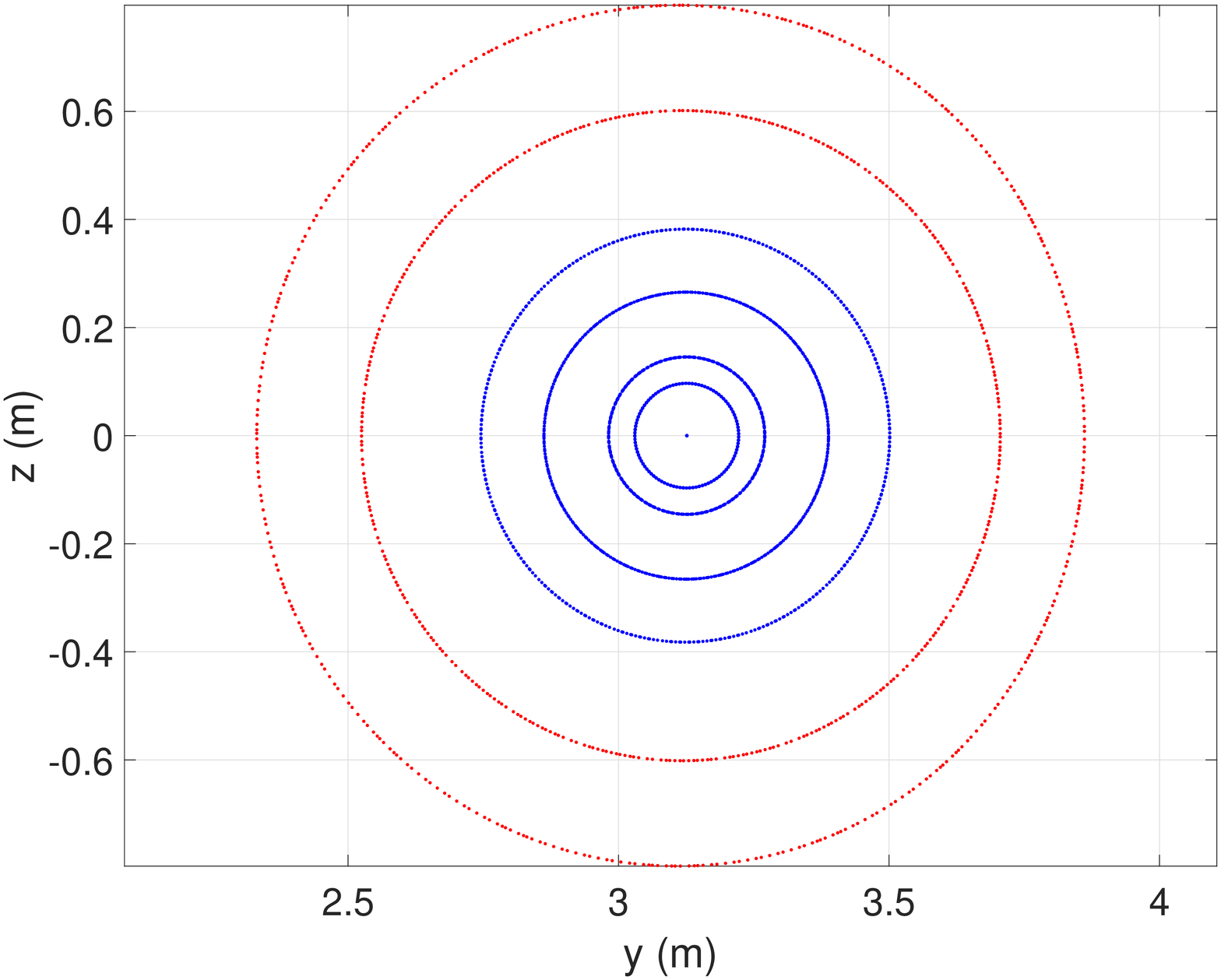}}
\caption{Cross section of magnetic flux surfaces of the linked mirror configuration, where $\phi$ is the toroidal angle}\label{fig2}
\end{figure}

\begin{figure}[h]
\includegraphics[scale=0.25]{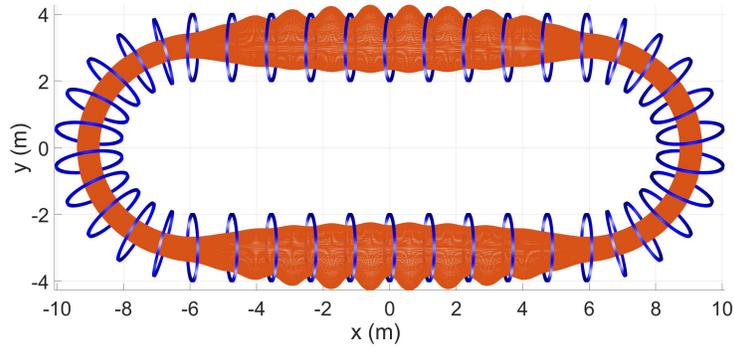}
\caption{The last closed magnetic flux surface with the coils}
\label{fig3}
\end{figure}

Fig. \ref{fig2} plots the cross-section of the magnetic flux surfaces of LMC obtained from field line tracing for the $\phi =0$ plane (a) and the $\phi=\pi/2$ plane (b) respectively. Here $\phi$ is the toroidal angle about the z-axis through the center of the whole configuration.
It shows that by this type of simple connections, nice magnetic flux surfaces are formed with finite rotational transform. The magnetic flux surfaces plotted in red indicate that the flux surfaces go out of the cylindrical tube within coils,
as shown in Fig. \ref{fig3}. They are still closed magnetic flux surfaces and can potentially be used as divertors.

\section{Rotational transform properties}

The LMC as just defined above is a special toroidal magnetic confinement configuration with finite external rotational transform generated by the coils. The mechanism for the rotational transform is the 3D magnetic field lines induced by finite $\Theta$ just as in the early figure-8 stellarator. In this section, we will show the dependence of the rotational transform on key configuration parameters.

\begin{figure}[h]
\includegraphics[scale=0.3]{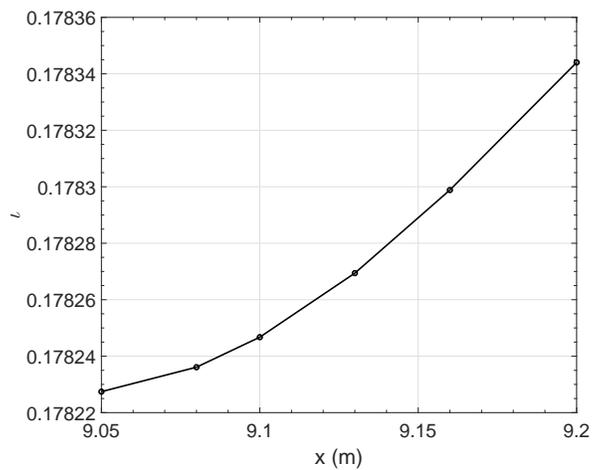}
\caption{Rotational transform versus $x$, where $x$ is the starting point of each field line with $z=0$ at $\phi=0$ plane}
\label{fig4}
\end{figure}

Fig. \ref{fig4} plots the rotational transform profile against radius variable $x$ for the LMC plotted in Fig. \ref{fig1},
where $x$ is the major radius at $\phi=0$ plane.
We observe that the rotational transform profile is almost constant with little magnetic shear.

\begin{figure}[h]
\subfigure[Rotational transform versus $\Theta$]{
\label{fig5.a}
\includegraphics[scale=0.2]{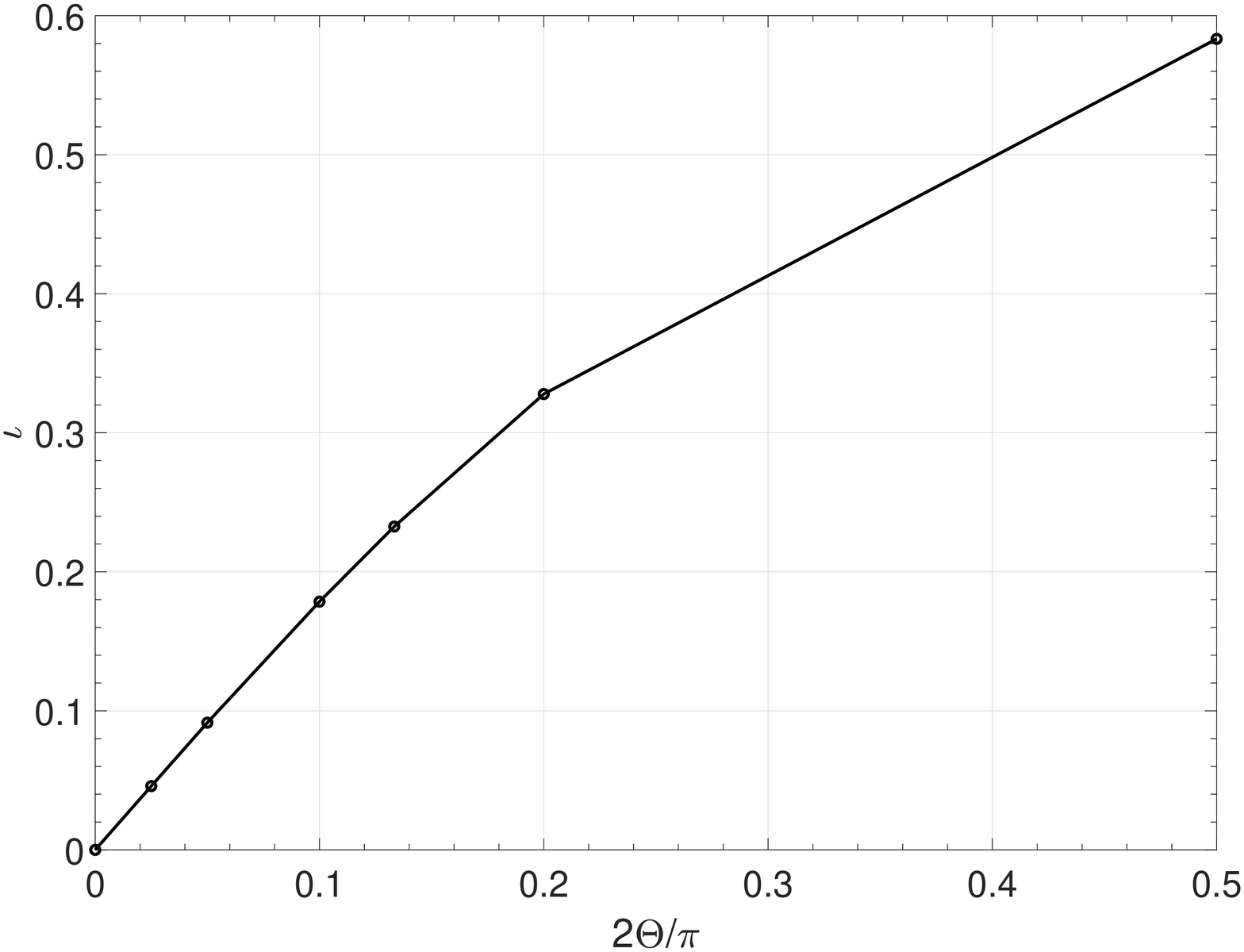}}
\subfigure[Rotational transform versus $L$]{
\label{fig5.b}
\includegraphics[scale=0.2]{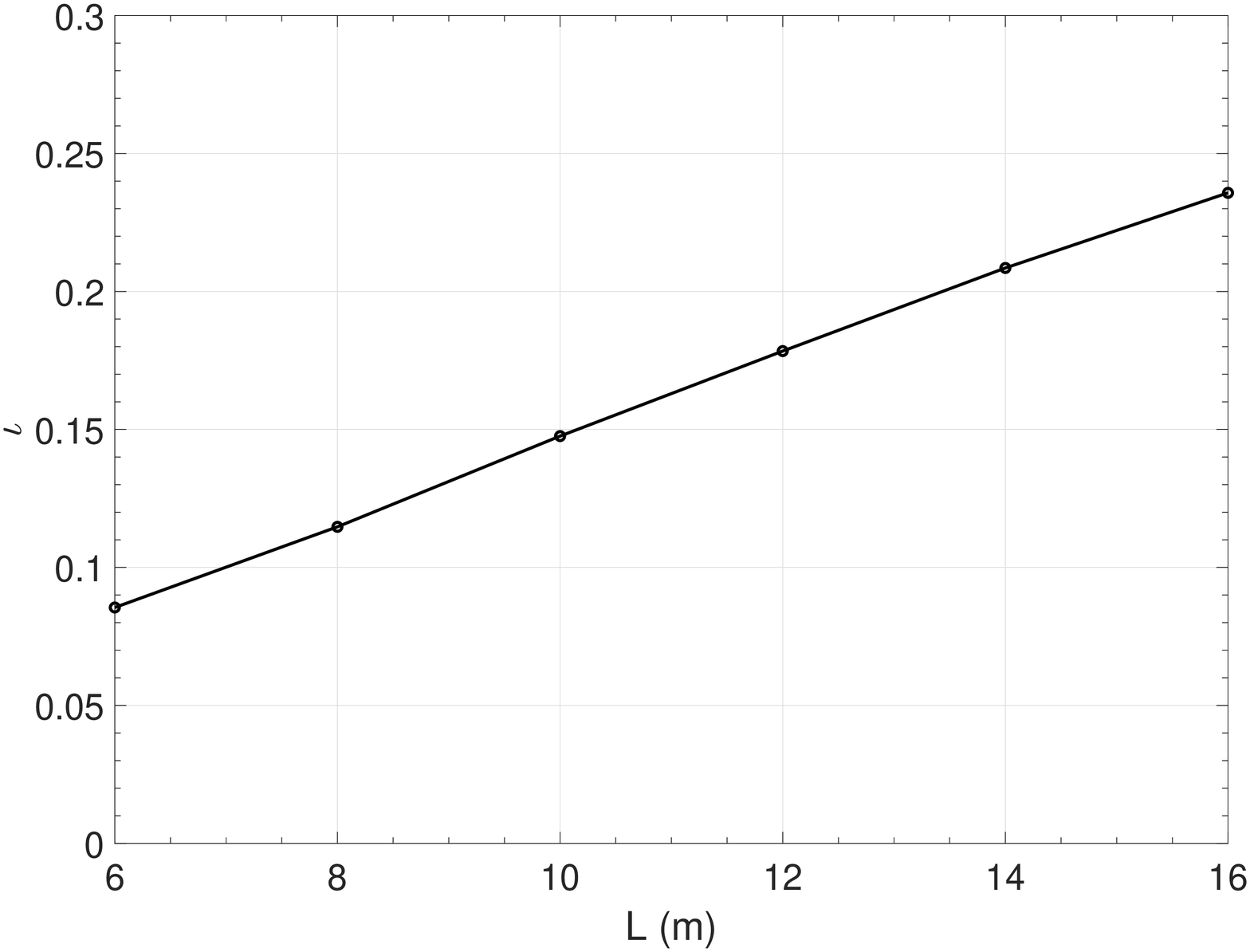}}
\subfigure[Rotational transform versus $d$]{
\label{fig5.c}
\includegraphics[scale=0.2]{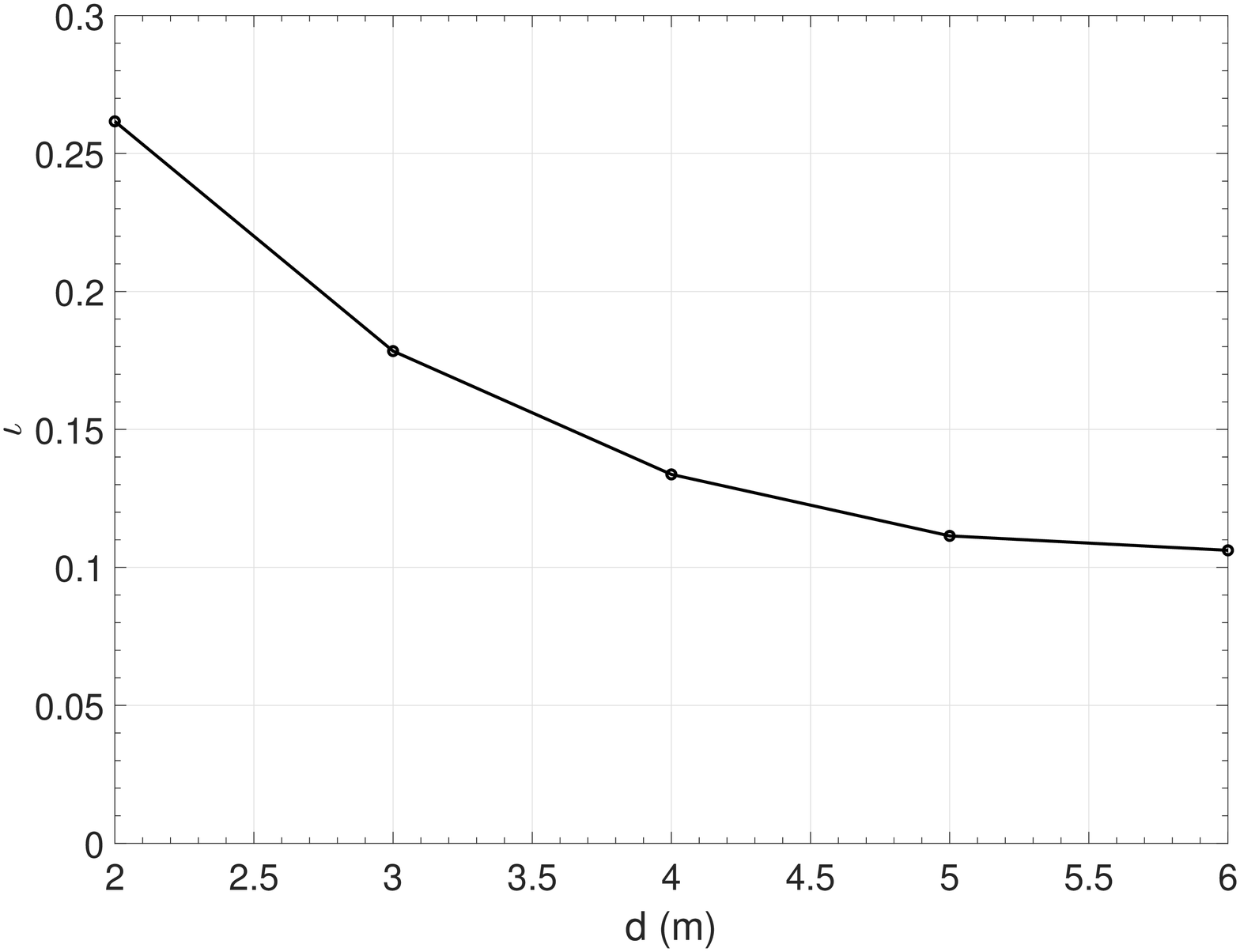}}
\caption{The dependence of rotational transform on the main parameters}\label{fig5}
\end{figure}

Fig. \ref{fig5.a}, \ref{fig5.b} and \ref{fig5.c} plot the dependence of the rotational transform on
$\Theta$, $L$ and $d$ respectively. For each plot, the following parameters are held fixed unless it is varied: $\Theta=\pi/20$, $L=12m$, $d=3m$, $r=1m$, $I_{tours}/I_{tube}=5I_0:1I_0$.  Fig. \ref{fig5.a} shows that the rotational transform of the LMC increases almost linearly with $\Theta$ and can vary in a large range, from 0 to 0.58. The last point in Fig. \ref{fig5.a} refers to the case where the two straight mirror tubes are perpendicular to each other. Configurations with $\Theta>\pi/4$ have higher rotational transforms. However, the flux surfaces would become too small relative to the coil size when the angle is larger than $45^\circ$. Furthermore
Fig. \ref{fig5.b} shows that the rotational transform increases with $L$, the length of the straight mirror sections. It should be noted that in the limit of $L=0$, the configuration becomes an axisymmetric torus with $\iota=0$ (since there is no plasma current). Finally, Fig. \ref{fig5.c} shows that the rotational transform decreases when $d$ increases, but it tends to a fixed value when $d\rightarrow\infty$.

In summary, the rotational transform profile of this linked mirror configuration is shearless, and can be varied in a vast range by simply changing one of the geometry parameters. This property can be used to avoid low order rational iota values in order to control MHD instabilities.

\section{Quasi-omnigeneity and single particle confinement}

In this section we consider the collisionless single particle confinement in the LMC. We will show that the magnetic configuration is quasi-Omnigeneous resulting in good neoclassical particle confinement. Single particle orbit simulations are carried out using the drift-kinetic equation with the magnetic field calculated from all coils. A numerical code with high accuracy has been developed for this purpose. The details of the code will be reported elsewhere.

\begin{figure}[h]
\subfigure[]{
\label{fig6.a}
\includegraphics[scale=0.2]{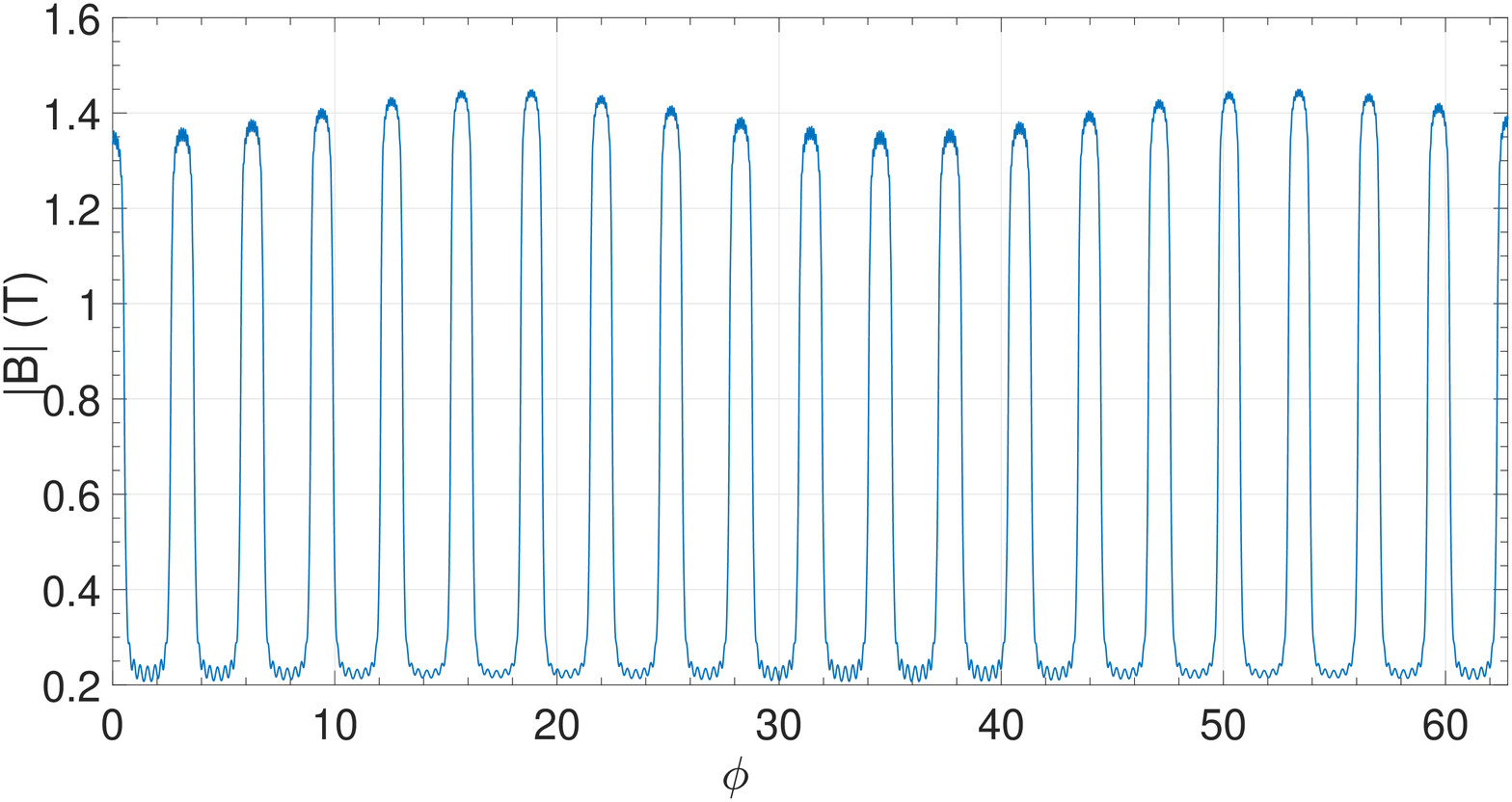}}
\subfigure[]{
\label{fig6.b}
\includegraphics[scale=0.2]{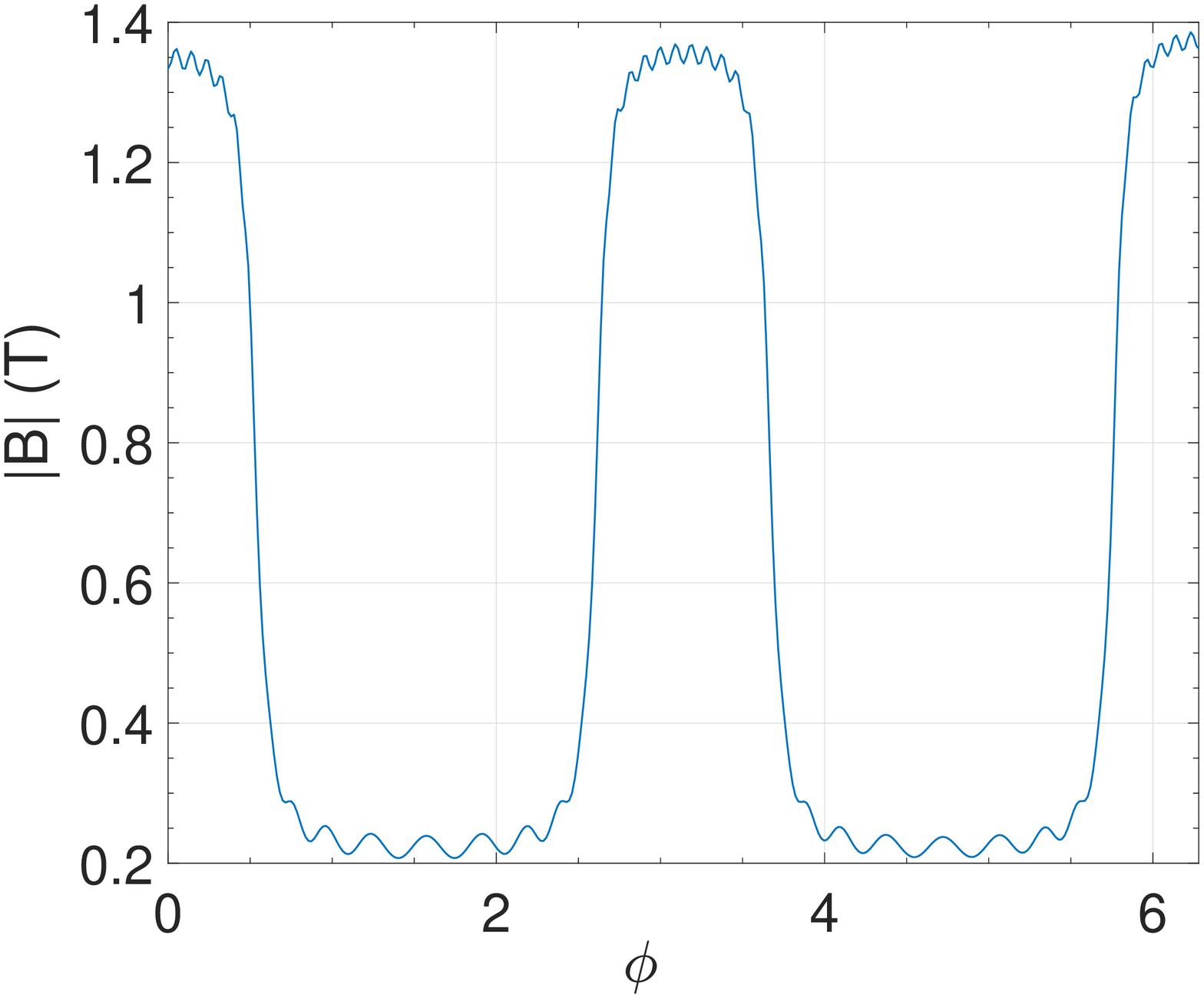}}
\caption{Magnetic field strength along the field line starting at $(9.15,0,0)$}
\label{fig6}
\end{figure}

Fig. \ref{fig6} plots the magnetic field strength along one field line starting at $(x,y,z)=(9.15,0,0)$ at the reference coil current $I_0=2\times10^5$A. The valleys are in the straight mirror sections, while the peaks are in the half-torus sections. The minimum values of the field strength along one field line are almost the same, and the maximum values also do not vary very much. This indicates that the magnetic configuration is quasi-omnigeneous\cite{matt}. There are ripples along the field line (see the expanded view in Fig. \ref{fig6.b}), which comes from the effect of discrete coils. Omnigeneity is defined by\cite{Laurence,Helander}:
\begin{equation}\label{eq_omnigeneity}
<\mathbf{v_d}\cdot \nabla\psi>_b=\int \mathbf{v_d}\cdot \nabla\psi dt=0
\end{equation}
for all trapped particles, where $\mathbf{v_d}$ is the guiding center drift velocity. In another word, the bounce-averaged drift velocity is zero or very small.

\begin{figure}[h]
\includegraphics[scale=0.3]{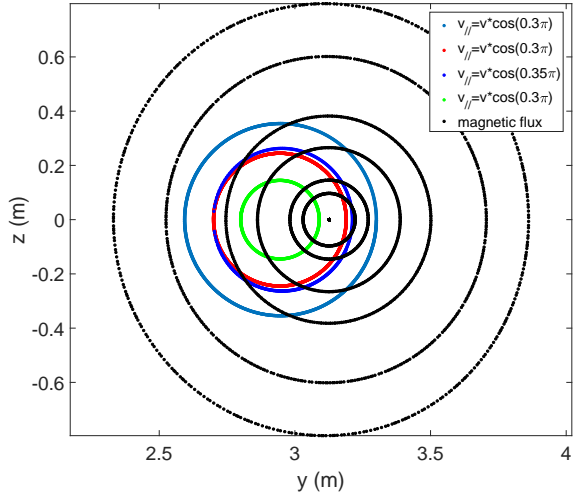}
\caption{Poincare plots of mirror trapped particles' orbits at $\phi=\pi/2$ plane, energy E=100eV}
\label{fig7}
\end{figure}

Fig. \ref{fig7} shows the Poincare plot for the orbits (colored lines) of mirror-trapped particles as well as flux surfaces (black lines) at the $\phi=\pi/2$ plane. We observe that these orbits are all closed. Furthermore, the poloidal processing frequency $\omega_d$ is much smaller than the bounce frequency $\omega_b$ ($\omega_b\approx500\omega_d$). These results indicate that $<\mathbf{v_d}\cdot \nabla\psi>_b$ is vary small, which means that the configuration is quasi-omnigeneous.

\begin{figure}[h]
\includegraphics[scale=0.3]{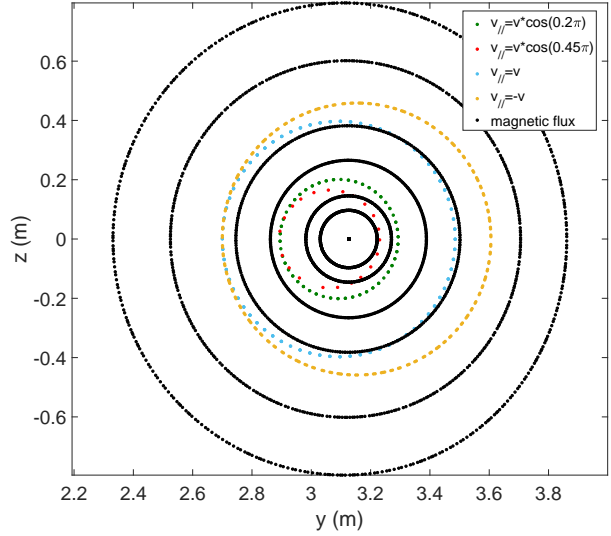}
\caption{Poincare plots of passing particles' orbits at $\phi=\pi/2$ plane, energy E=100eV}
\label{fig8}
\end{figure}

There exist finite shifts between the trapped particles' orbits and the magnetic flux surfaces. This is due to the fact that the magnetic flux surfaces are not cylindrically symmetric as the coils in the mirror sections because the configuration is 3D. In particular we observe that the position of magnetic axis is not at the center of coils as shown in Fig. \ref{fig1} and Fig. \ref{fig3}. This orbital shift can be reduced to zero by careful placement of coil locations in the half-torus sections. Furthermore we observe that, just like trapped particles, passing particles' orbits also have finite radial shift away from the flux surfaces as shown in Fig. \ref{fig8},  although the size of orbital shifts is usually quite modest.

\begin{figure}[h]

\includegraphics[scale=0.35]{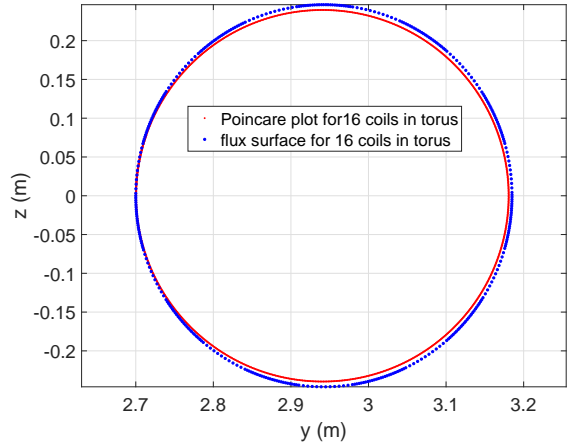}
\caption{Reference flux surface and Poincare plot for particle starting at same point (0,2.7,0), with energy E=100ev and pitch angle ($v_{||}=v\cos(0.3\pi)$)}
\label{fig8_add}
\end{figure}

We found that by placing different numbers of coils in the half-torus sections, the flux surfaces in straight section would move translationally, while the trapped particle's orbits keep almost fixed. In particular, the configuration with 16 coils in half-torus sections aligns the magnetic surfaces and the poloidally processing orbits of trapped particles better than other configurations(Fig. \ref{fig8_add}).

\begin{figure}[h]
\includegraphics[scale=0.3]{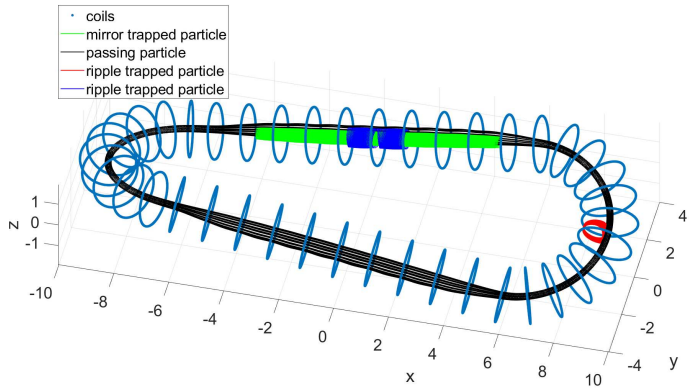}
\caption{Orbits for the typical passing particles, mirror trapped particles and ripple trapped particles}
\label{fig9}
\end{figure}

 A systematic study of single particle confinement has been done. Simulation results show that
there are four types of particle orbits in the LMC: passing particles, mirror trapped particles, ripple trapped particles, and particles trapped by the magnetic field variation with higher magnetic field strength mainly in the half-torus sections(particles with $E/\mu\sim1.36-1.45T$ in Fig. \ref{fig6}). These "trapped particles" are in fact passing particles processing toroidally. However, their parallel velocity changes sign after several toroidal transits. Passing particles and mirror trapped particles are found to be confined well. Ripple trapped particles are also confined in the configuration although they may leave the last closed surface. Fig. \ref{fig9} plots the orbits for the typical passing particles, mirror trapped particles and ripple trapped particles.

The particles barely trapped by the magnetic field in the mirror sections may not be confined for a long time, since they bounce between two points with high fields near the half-torus sections in one period and the magnetic drift due to the toroidal curvature is not averaged out. The fraction of these lost particles is small since they are barely trapped.

\begin{figure}[h]
\subfigure[Number of left particles versus t for 10000 test particles with energy E = 1keV]{
\label{fig10.a}
\includegraphics[scale=0.25]{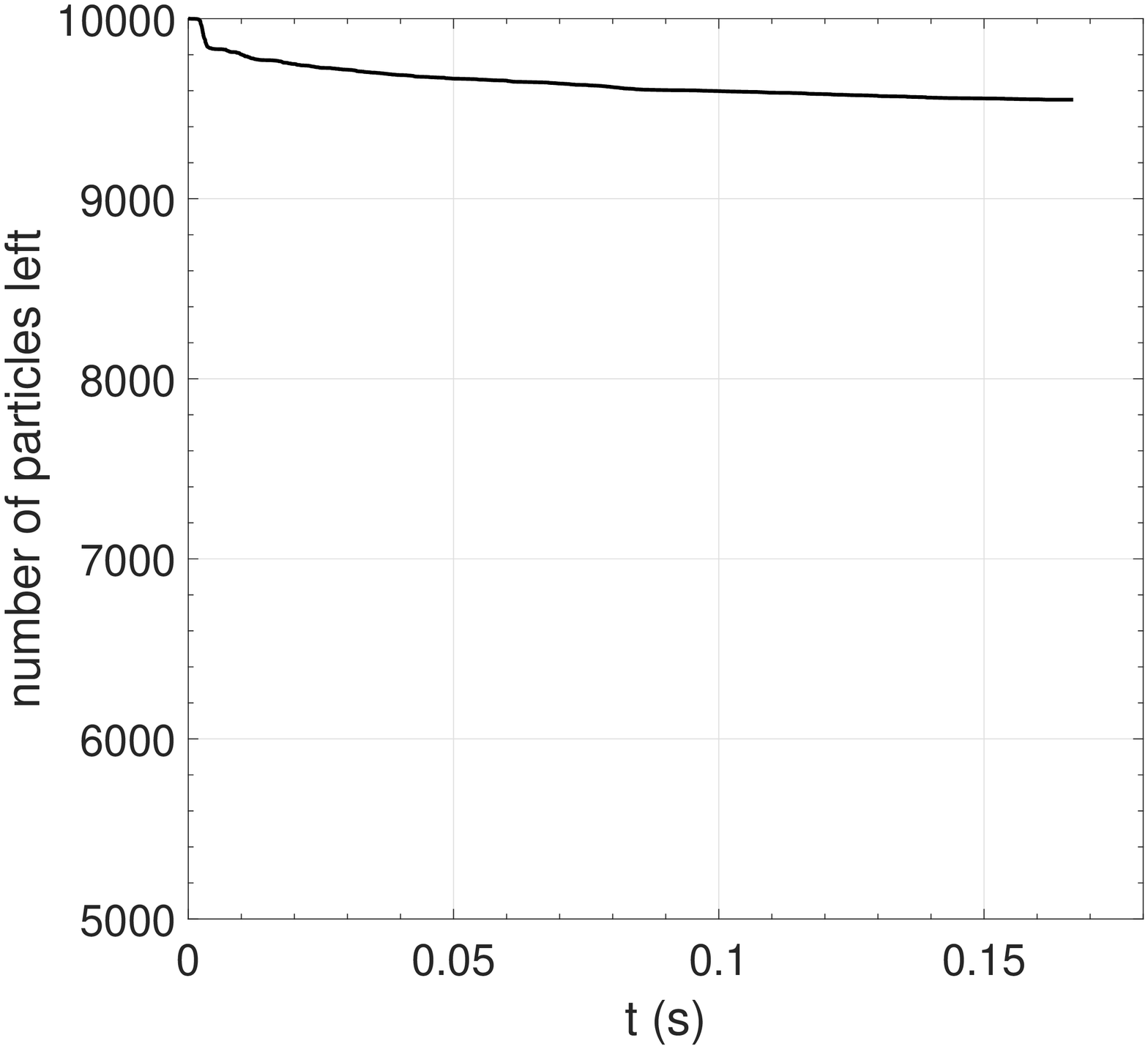}}
\subfigure[Distribution of test particles and loss particles with energy E = 1keV in $v_{\parallel}/v$ and $\phi$ space]{
\label{fig10.c}
\includegraphics[scale=0.25]{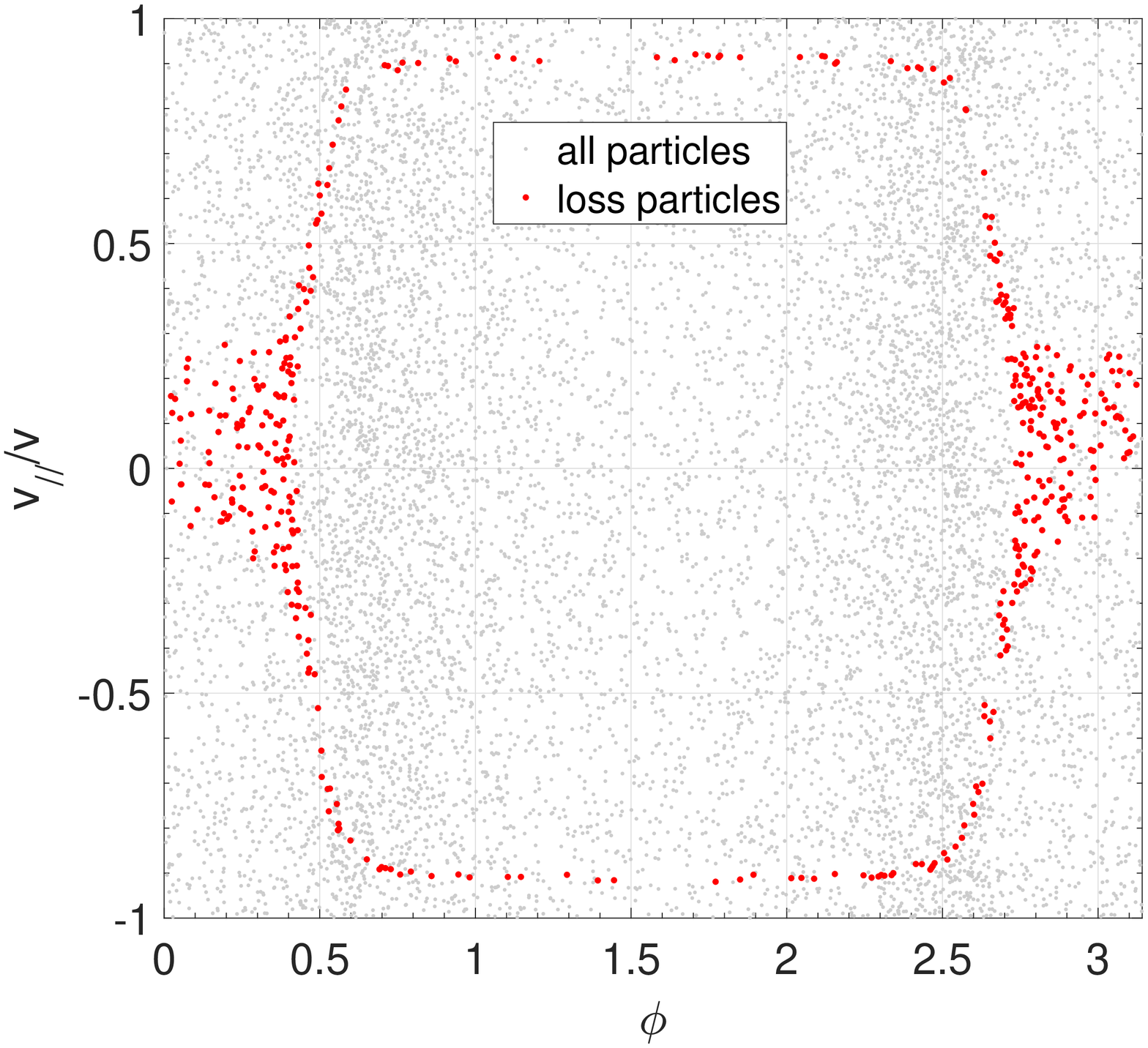}}
\subfigure[Distribution of test particles and loss particles with energy E = 1keV in $\mu B_{max}/E$ and $\phi$ space]{ \label{fig10.d}
\includegraphics[scale=0.25]{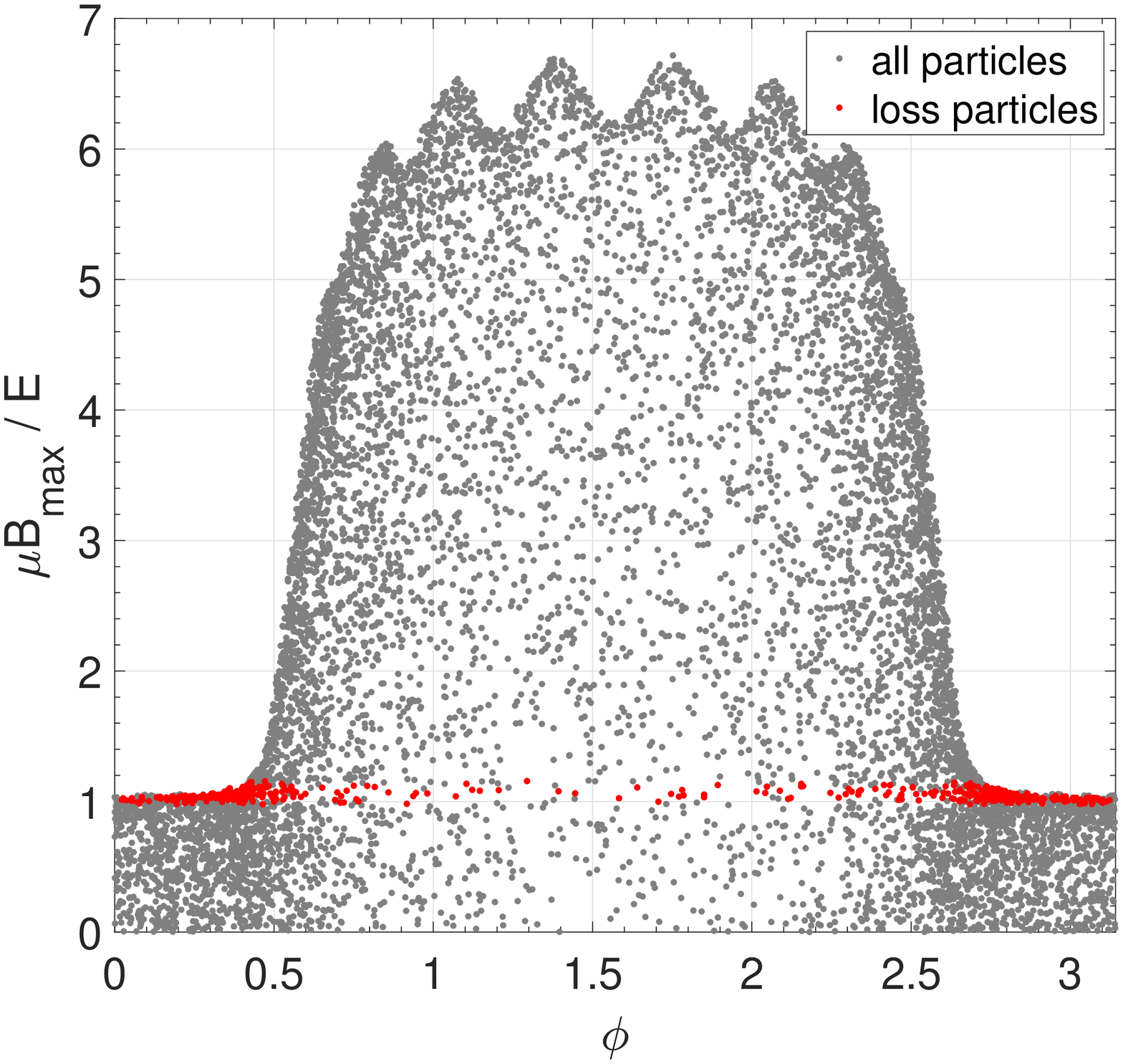}}
\caption{Study of 10000 test particles energy E = 1keV in the linked mirror configuration}\label{fig10}
\end{figure}

\begin{figure}[h]
\includegraphics[scale=0.3]{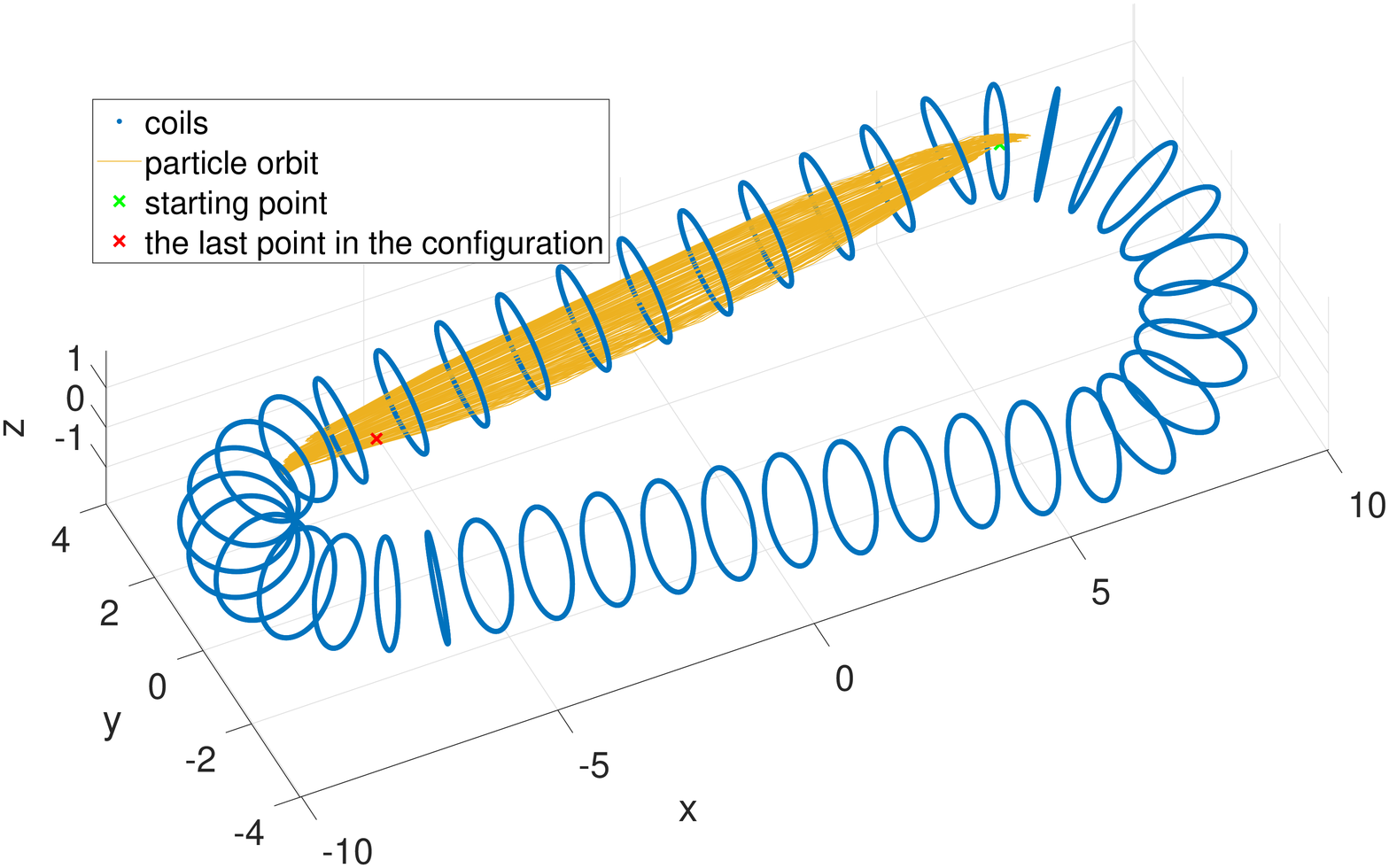}
\caption{Orbit of the typical loss particles}
\label{fig11}
\end{figure}

Fig. \ref{fig10} plots the simulation results of 10000 test particles with energy E = 1keV in the LMC given in Fig. \ref{fig1}. The particles are initially distributed uniformly in real space on the middle magnetic flux surface going through the point (9.1,0,0). The velocity distribution is isotropic. The particle boundary is chosen to be at $r_p=0.99r$, where $r$ is the coil radius. Fig. \ref{fig10.a} shows that more than $95\%$ particles are still confined after a long duration (about 0.17s). There is a thin envelope formed by loss particles in phase space as shown in Fig. \ref{fig10.c}.  These lost particles are barely trapped particles near the trap-passing boundary in the mirror sections. All the passing particles are well confined including those "trapped" by the global magnetic well at high fields. There is another type of particles: $v_{//}/v\sim0$, which are confined by ripple wells in both mirror sections and the half-torus sections. They are well confined in the configuration as shown in Fig. \ref{fig9}. We observe in Fig. \ref{fig10.d} that all the lost particles have similar $\mu B_{max}/E$ values ($\approx1$). This indicates that the lost particles in the LMC are all near the passing-trap boundary in phase space. As a result the loss cone is very narrow. Furthermore,  unlike first orbit loss in tokamaks, these particles typically bounce back and forth for tens to hundreds times before leaving the configuration as shown for one such orbit in Fig. \ref{fig11}.

\begin{figure}[h]
\subfigure[Fraction of confined particles versus time for particles with different energies]{
\label{fig12.a}
\includegraphics[scale=0.25]{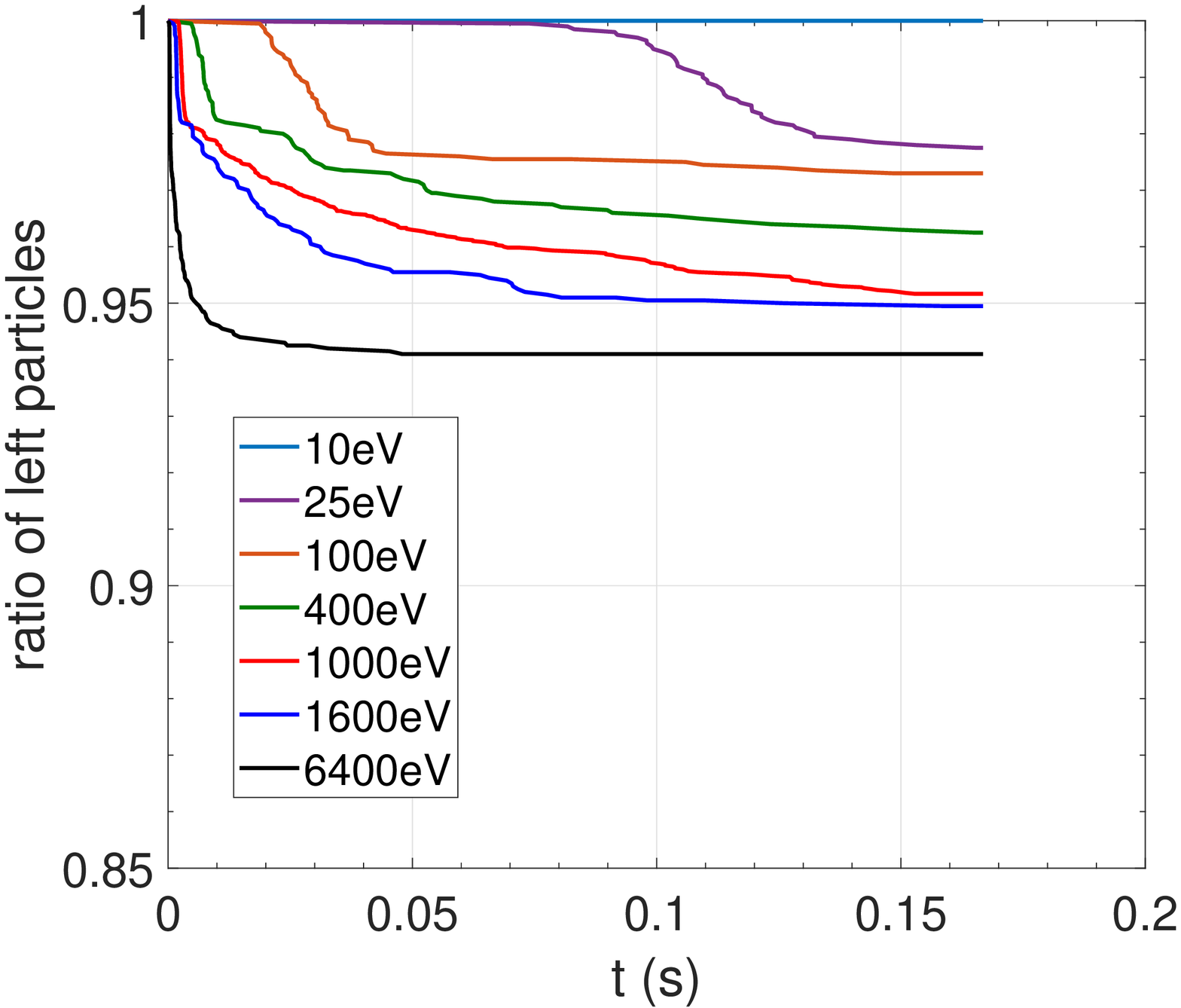}}
\subfigure[Fraction of confined particles versus particle energy after 0.1668s]{
\label{fig12.b}
\includegraphics[scale=0.25]{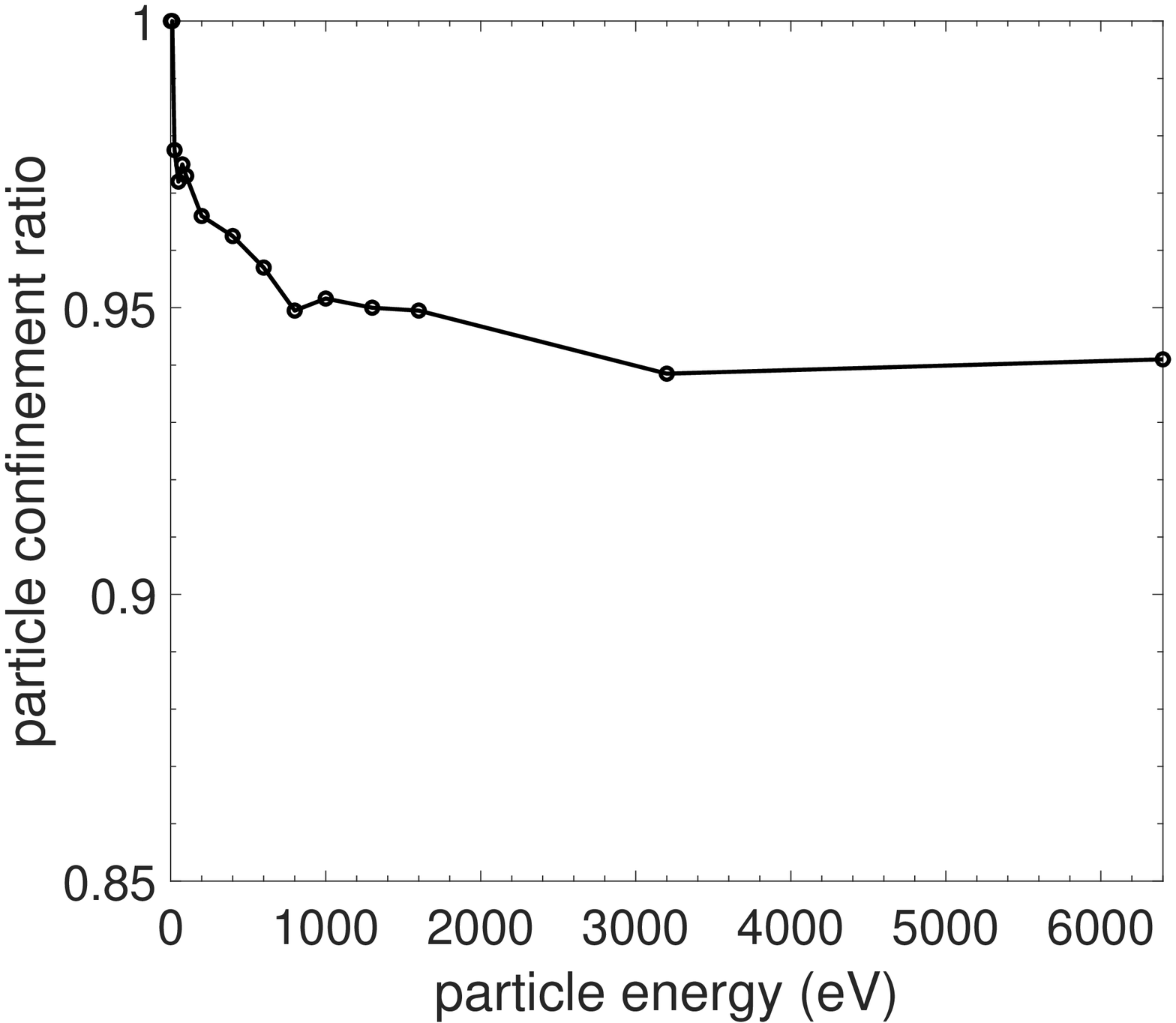}}
\caption{Particle confinement scaling study}\label{fig12}
\end{figure}

Fig. \ref{fig12.b} shows that the fraction of confined particles tends to a constant after a long time ( 0.1668s ) for particle energy larger than 1600eV. It is a sufficiently long time period to calculate the fraction because the number of confined particles becomes almost a constant at that time, as shown in Fig. \ref{fig12.a}.

\section{Neoclassical transport}

\begin{figure}[h]
\includegraphics[scale=0.3]{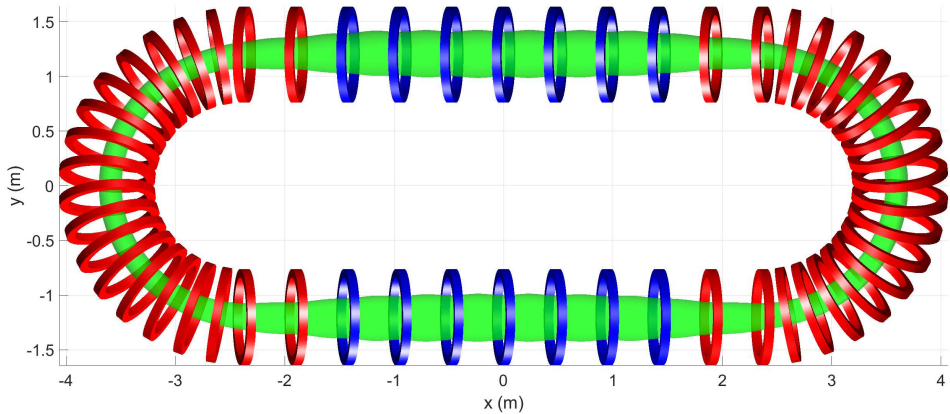}
\caption{Overview of the linked mirror with 54 coils}
\label{fig13}
\end{figure}

We have pointed out that the trapped particle's bounce orbit width is very small as compared with the device size and the neoclassical transport level is expected to be very low. Here we investigate the neoclassical transport numerically. The particle collisions are taken into account using the Monte Carlo pitch angle scattering model\cite{Boozer}.  We consider a linked mirror configuration with 54 coils as shown in Fig. \ref{fig13}. The following key parameters are used: $\Theta=\pi/20$, $L=4.8m$, $d=1.2m$, $r=0.4m$, $I_{red}/I_{blue}=160kA:80kA$. The corresponding mirror ratio is about $R=B_{max}/B_{min} = 0.87T/0.22T \sim 4$. It should be pointed out that the plasma minor radius ($a_0=0.2m$) is about $50\%$ of the size given in Fig. 1. More coils are used here because they help to align the magnetic axis with the coil center in the straight mirror sections. As a result, the precession orbits of mirror-trapped particles(Fig. \ref{fig7}) almost coincide with magnetic flux surfaces.

\begin{figure}[h]
\subfigure[Fraction of confined particles versus time for T=160eV protons]{
\label{fig14.a}
\includegraphics[scale=0.25]{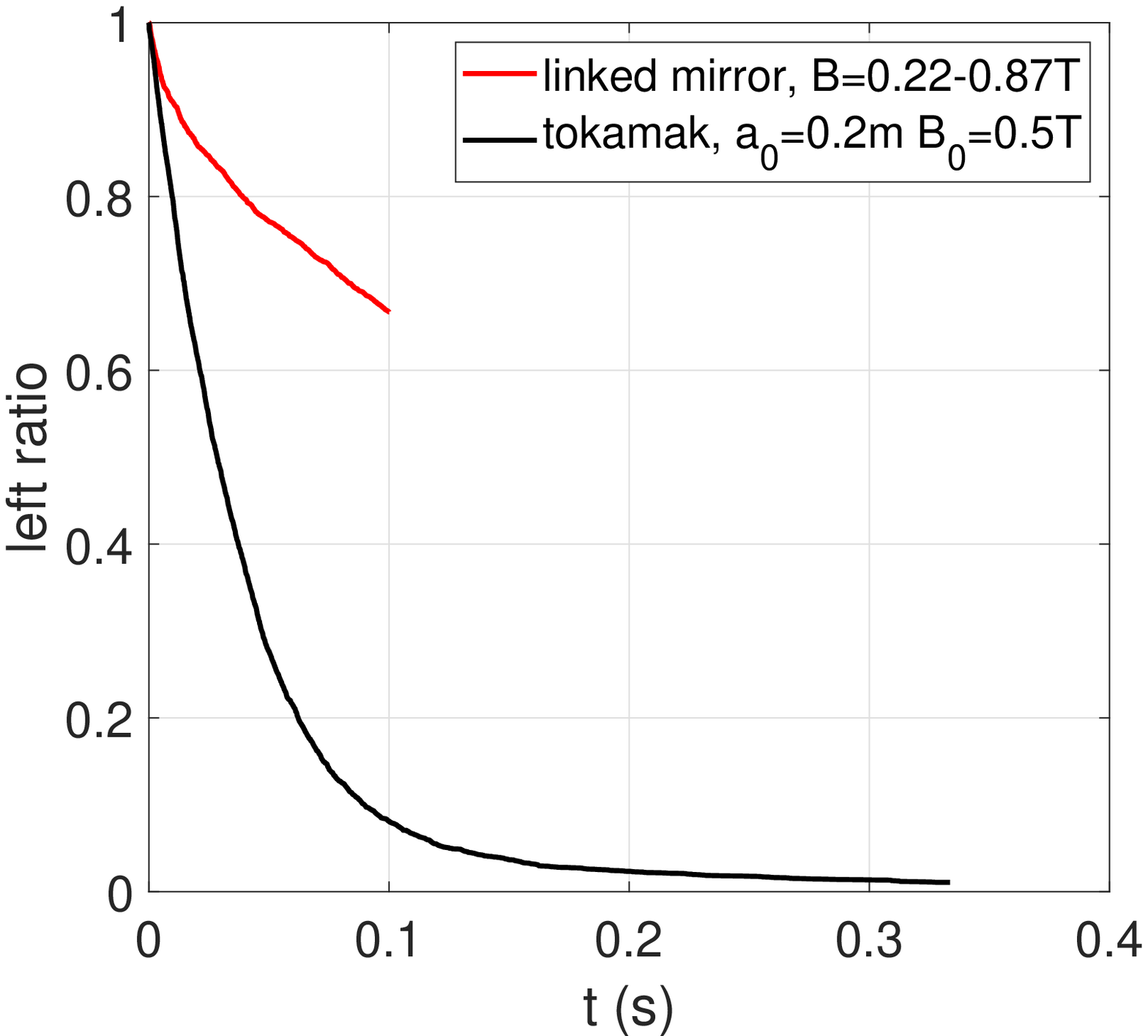}}
\subfigure[Real time loss rate for T=160eV protons]{
\label{fig14.b}
\includegraphics[scale=0.25]{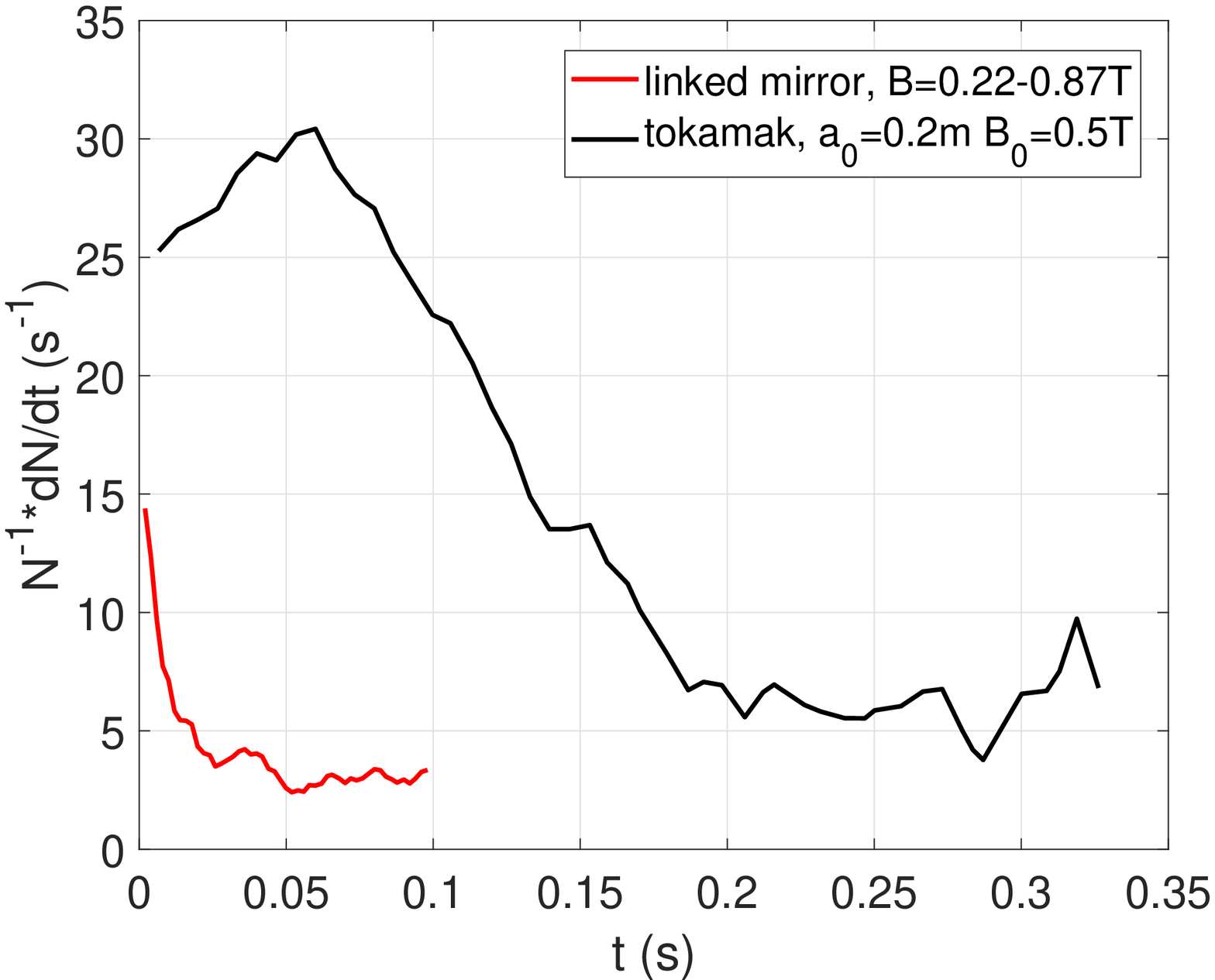}}
\caption{Neoclassical transport study for linked mirror configuration and tokamak. The collision rate is proportional to $E^{-3/2}$, and the same for protons with same energy in both configuration.}\label{fig14}
\end{figure}

Fig. \ref{fig14} plots the numerical results of neoclassical transport studies. At the beginning of particle simulations, 4000 protons are loaded uniformly on the middle flux surface. The velocity distribution is a Maxwellian with temperature $T=160eV$. The plasma density is chosen to be $6.4\times10^{17}m^{-3}$. Fig. \ref{fig14} shows the fraction of confined ions as a function of time (a) and the real time loss rate (b) for the new configuration (red line) and an equivalent tokamak (black line). Here the real time loss rate is estimated by $N^{-1}dN/dt$. It can be regarded as the inverse of particle confinement time if the loss rate converges to a steady state value in time.
The results clearly show that the particle confinement of the considered configuration is substantially better than that of an equivalent tokamak. We find that the particle confinement time for the configuration is about 0.3s, which is quite large considering the parameters used.

\section{MHD stability}

So far we have only considered the property of vacuum magnetic field with focus on particle confinement. Concerning MHD stability, our configuration at zero or small plasma beta has a small magnetic hill on order of $1\%$, i.e., the strength of flux surface-averaged magnetic field decreases about $1\%$ from magnetic axis to the last closed flux surface. As a consequence the interchange mode is expected to be unstable, at least for small plasma beta assuming pressure is constant on magnetic surfaces. There are a number of ways to stabilize
this interchange mode\cite{Ryutov} including stabilization by energetic sloshing ions\cite{Hinton} and Finite ion Larmor Radius (FLR) effects\cite{Rosenbluth}.
First, the instability can be mitigated or even stabilized by choosing a value of rotational transform such that the low order rational resonances are avoided since the rotational transform profile is basically uniform. Second, a simple estimate
has shown that the interchange mode can be stabilized by energetic sloshing ions with a peak pitch angle distribution (i.e., via neutral beam injection) if energetic ion pressure is comparable to background plasma pressure. Finally the FLR stabilizing is expected to be significant since the poloidal mode numbers of the interchange modes are large (note that the value of rotational transform is small).
Therefore our linked mirror configuration as designed is potentially MHD stable at high beta on order of unity. A more definitive conclusion requires calculation of finite beta equilibria and a more precise determination of MHD stability. This will be done in a future work.

\section{Conclusions}

In conclusion, a linked mirror configuration is proposed. The new device consists of two linear mirror sections with ends connected by half-torus sections. The magnetic field is generated by simple circular planar coils. The passing particles are well confined due to rotational transform generated by a finite angle between the two linear mirror sections. Therefore the usual end loss of linear mirror devices is completely eliminated.
Most particles are well confined including the mirror trapped particles and the ripple-trapped particles. Monte Carlo simulations show that the neoclassical confinement is very good and is even better than that of an equivalent tokamak. The proposed linked mirror configuration is potentially suitable for neutron sources and fusion reactors.

\begin{acknowledgments}
This work is supported by Zhejiang University's startup funding for one of the authors (Prof. Guoyong Fu).
\end{acknowledgments}

\end{document}